# Liquid-core low-refractive-index-contrast Bragg fiber sensor

Hang Qu, Maksim Skorobogatiy*

*Genie physique, Ecole Polytechnique de Montrea, C.P.6079, succ. Centre-ville, Montreal, QC, Canada, H3C 3A7l*

*Corresponding author: maksim.skorobogatiy@polymtl.ca

We propose and experimentally demonstrate a low-refractive-index-contrast hollow-core Bragg fiber sensor for liquid analyte refractive index detection. The sensor operates using a resonant sensing principle— when the refractive index of a liquid analyte in the fiber core changes, the resonant confinement of the fiber guided mode will also change, leading to both the spectral shifts and intensity changes in fiber transmission. As a demonstration, we characterize the Bragg fiber sensor using a set of NaCl solutions with different concentrations. Strong spectral shifts are obtained with the sensor experimental sensitivity found to be ~1400nm/RIU (refractive index unit). Besides, using theoretical modeling we show that low-refractive-index-contrast Bragg fibers are more suitable for liquid-analyte sensing applications than their high-refractive-index-contrast counterparts.

Recently, liquid-core waveguides have drawn much attention for chemical and biological sensing applications. A typical liquid-core waveguide sensor relies on the principle of total internal reflection (TIR) where liquid-core is surrounded by the lower refractive index (RI) cladding. Unfortunately, the difficulty in finding suitable cladding materials whose refractive indices are lower than those of aqueous solutions (n~1.33), limits TIR liquid-core sensor development. One way to circumvent this problem is to use the "leaky modes" guided in the low-RI core surrounded by the high-RI cladding. This is an approach used by the liquid- and air- core capillary sensors reported in [1-3]. Although easy to fabricate, these capillary sensors, however, have limited sensing length due to the large propagation loss, thus requiring additional process such as deposition of reflective metallic coatings [1, 2].

In this paper, we propose and experimentally demonstrate a liquid-core fiber sensor using the low-RI-contrast hollow-core (HC) Bragg fiber fabricated in our group [4]. The Bragg fiber features a large air core (diameter: 0.5-1mm) surrounded by an alternating polymethyl methacrylate (PMMA) / polystyrene (PS) Bragg reflector (RI: 1.49/1.59 @589nm) and a PMMA cladding as shown in figure 1. Unlike the above-mentioned capillary sensors, Bragg fibers guide by bandgap effect – the light within a certain frequency (bandgap) is confined to the fiber core due to strong reflection from a periodic reflector. The propagation loss of guided light is then determined by absorption loss of a core material and fiber radiation loss. Moreover, the fiber can be coiled into a several meter long sensing cell with virtually no additional loss. The large core of a Bragg fiber facilitates filling it with aqueous solutions that might even contain large objects such as bacteria clusters. The response time of the sensor is ~ 1s. The Bragg fiber sensor proposed in this paper operates using a resonant sensing principle according to which the fiber transmission spectrum changes in response to the changes in the real part of the analyte RI filling the core [5].

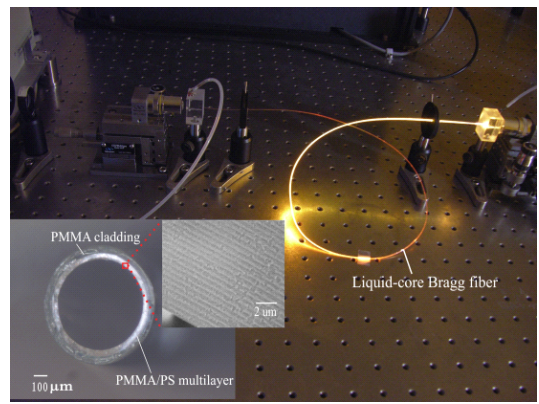

Figure 1. Experimental setup of the liquid-core Bragg fiber sensor; the 80cm fiber is coiled into a 20cm-diameter coil. The inset is cross section of the Bragg fiber.

From the basic theory [6] of the low-RI-contrast Bragg fiber, the center wavelength, $\lambda_c$, of a reflector bandgap is approximately given by:

$$\frac{\lambda_c}{2} = d_h(n_h^2 - n_c^2)^{1/2} + d_l(n_l^2 - n_c^2)^{1/2} \qquad (1)$$

where $d_l$, $d_h$ are the thicknesses of the low- and high-RI layer respectively; $n_l$, $n_h$ are the real part of refractive indices of the corresponding layers; $n_c$ is the real part of the RI of the core material. If the real part of the analyte RI changes, the resonant condition (1) of the guided modes will also be modified, leading to the changes in the center position of fiber bandgaps, which is the main operation principle of this fiber sensor. We note that low-RI-contrast Bragg fibers have certain advantages for liquid analyte sensing, compared to their high-RI-contrast counterparts. The high-RI-contrast Bragg fiber has shown superior air-core (n~1) guidance which was used for gas sensing application [7]; K. J. Rowland, *et al* also reported guiding and sensing properties of the high-RI-contrast Bragg fiber filled with high RI liquids (n>1.40) [8]. However, for most high-RI-contrast Bragg fibers with aqueous solutions (n~1.33) filled in the core, the TM bandgaps of the Bragg reflector tend to collapse at the light line of aqueous material due to the Brewster angle phenomenon, leading to high loss for the hybrid (HE/EH) modes. In contrast, the low-RI-contrast Bragg fibers show large TM bandgaps in the vicinity of the light line of the aqueous core, thus resulting in good guidance. Moreover, from equation (1) we derive the Bragg fiber sensor sensitivity, $S$, as:

$$S = \left| \frac{d\lambda_c}{dn_c} \right| = 2[d_h(\frac{n_h^2}{n_c^2} - 1)^{-1/2} + d_l(\frac{n_l^2}{n_c^2} - 1)^{-1/2}] \qquad (2)$$

According to equation (2), the closer is the value of the core RI to those of the individual layers of the Bragg reflector, the more sensitive the Bragg fiber sensor will be to the changes in the core RI which is exactly the case for the low-RI-contrast Bragg fibers.

To verify the sensing principle theoretically, we simulate the loss spectra of the fundamental mode ($HE_{11}$ mode) of the liquid Bragg fiber based on the Transmission Matrix Method (TMM). The average thicknesses of the high- and low- index layer, obtained from the SEM graphs, are 0.13μm and 0.37μm respectively; the number of bilayers in the Bragg reflector is approximately 25; the refractive indices of PMMA and PS films are measured by a spectroscopic ellipsometer. We choose, as liquid analytes, a set of NaCl solutions with the weight concentration ranging from 0 to 25% with a 5% increment step. The refractive indices of NaCl solutions are obtained from [9]. The bulk absorption of NaCl solutions in the measuring range is assumed to be identical to that of water [10]. Our simulation suggests that the transmission spectrum shows a blue-shift as the increased RIs of NaCl solutions.

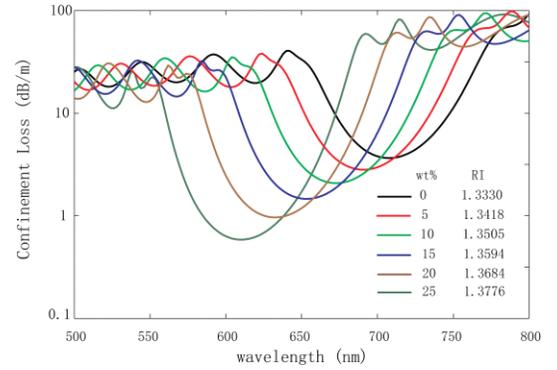

Figure 2. Simulated confinement loss of the fundamental $HE_{11}$ mode of the Bragg fiber filled with a set of NaCl solutions. The weight concentrations (wt%) and RIs of the NaCl solutions are listed in the figure.

Experimentally, we employ two microfluidic blocks to integrate a Bragg fiber into an optic-fluidic system. The two blocks are used to submerge both Bragg fiber tips "under water" to avoid air bubble infiltration which would strongly suppress the fiber transmission. The optical access to the fiber tip is through a lateral attached glass window on each block. The optical attenuation of in- and out- coupling is almost negligible. After filling a liquid sample into the sensor system, we couple a supercontinuum beam to the fiber (see, Fig. 1). The transmission spectra of the Bragg fiber filled with NaCl solutions with different concentrations are analyzed by a spectrometer (Fig. 3).

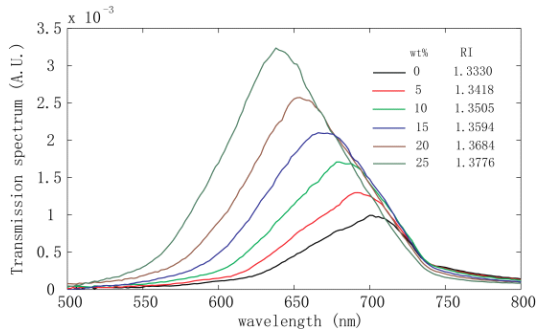

Figure 3. Experimental transmission spectra of Bragg fiber sensor filled with NaCl solutions.

In figure 3, we observe a blue-shift in transmission spectra of liquid Bragg fiber as the RI of the analyte increases, which agrees with the simulation result in figure 2. In figure 4, we show that the spectrum shift has a linear dependence to increment of analyte refractive index as can be seen both from the simulated and the experimental results.

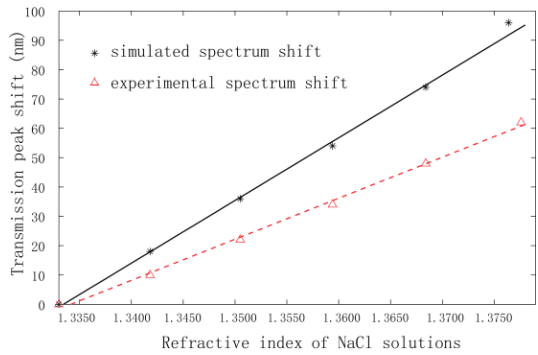

Figure 4. Transmission peak shift obtained from the experimental measurements (red dash line) and the TMM simulations (black solid line).

We notice that the experimental spectral shifts are somewhat smaller than those in the simulation. This is because in the simulation we only calculate the transmission of the fundamental mode ($HE_{11}$ mode), whose effective RI is closest in value to the RIs of the Bragg reflector material, which, according to equation (2), makes the $HE_{11}$ mode to be most sensitive to the changes in the core RI compared to other "m=1" modes. In practice, due to the large diameter of the Bragg fiber core, many high-order modes which are less sensitive to the core RI variations are also excited, resulting in smaller spectral shifts. The experimental sensitivity of the proposed sensor is ~1400nm/RIU, comparable to that of the surface plasmon resonance (SPR) sensor [11].

In conclusion, we report a low-RI-contrast hollow-core Bragg fiber sensor operating on a resonant sensing mechanism according to which the transmission spectrum shifts in response to the changes of the analyte RI. Both the theoretical simulations based on the TMM and the experiments were carried out to characterize the sensor. A sensitivity of ~1400nm/RIU is experimentally achieved, which is comparable to that of a SPR sensor. The proposed liquid Bragg fiber sensor configuration constitute an one-fiber solution for refractive index sensing with the advantage of the simplicity in sensing mechanism, large core, easy fabrication and high sensitivity.